\documentclass[floatfix,twocolumn,preprintnumbers,amsmath,amssymb,prb]{revtex4}
\usepackage{amsmath}
\usepackage{graphicx}%
\usepackage{dcolumn}
\usepackage{amsmath}
\usepackage{subfigure,hyperref}
\usepackage[normalem,normalbf]{ulem}

\usepackage{enumerate}

\newcommand{\be}{\begin{equation}}
\newcommand{\ee}{\end{equation}}

\usepackage{enumerate}

\def\nbR{\ensuremath{\mathrm{I\!R}}}

\begin{document}

\title{A hint of renormalization}
\author{Bertrand Delamotte}
\email{delamotte@lpthe.jussieu.fr}
\affiliation{Laboratoire de Physique Th\'eorique et 
Hautes Energies. Universit\'es Paris VI, Pierre et Marie 
Curie, Paris VII, Denis Diderot, 2 Place Jussieu, 75251 Paris
Cedex 05, France}

\begin{abstract}
An elementary introduction to perturbative renormalization and
renormalization group is presented. No prior knowledge of field
theory is necessary because we do not refer to a particular
physical theory. We are thus able to
disentangle what is specific to field theory and what is intrinsic
to renormalization. We link the general arguments and results to
real phenomena encountered in particle physics and statistical
mechanics.
\end{abstract}

\maketitle

\section{Introduction}
Hans Bethe in a seminal 1947 paper was
the first to calculate the energy gap, known as the Lamb shift, 
between the 2$s$ and 2$p$ levels of the hydrogen
atom.\cite{bethe47} These levels were found to be degenerate even
in Dirac's theory which includes relativistic corrections. Several
authors had suggested that the origin of the shift could be the
interaction of the electron with its own radiation field (and not
only with the Coulomb field). However, to quote Bethe, ``This
shift comes out infinite in all existing theories and has
therefore always been ignored.'' Bethe's calculation was the first
to lead to a finite, accurate result. Renormalization --- in its
modern perturbative sense --- was born.\cite{weinberg95} Since
then it has developed into a general algorithm to get rid of
infinities that appear at each order of perturbation theory in
(almost) all quantum Field theories
(QFT).\cite{collins84,ryder85,lebellac91,binney92,goldenfeld92} In
the meantime, the physical origin of these divergences has also
been explained (see Ref.~\onlinecite{cao99} for many interesting
contributions on the history and philosophy of renormalization
and renormalization group). 

In QFT, as in ordinary quantum mechanics, 
the perturbative calculation of any physical process involves, at
each order, a summation over (virtual) intermediate states.
However, if the theory is Lorentz invariant, an infinite number
of supplementary states exist compared with the Galilelan case
and their summation, being generically divergent,
produces infinities. The origin of these
``new'' states is deeply rooted in quantum mechanics and 
special relativity. When these two theories are combined, a new
length scale appears, built out of the mass $m$ of the particles:
the Compton wave length
$\hbar/mc$. It vanishes in both formal limits
$\hbar=0$ and $c=\infty$, corresponding respectively to classical
and Galilean theories. Because of Heisenberg inequalities,
probing distances smaller than this length scale requires energies
higher than $m c^2$ and thus implies the creation of particles.
This possibility to create and annihilate particles forbids the
localization of the original particle better than the Compton wave
length because the particles that have just been created are
strictly identical to the original one. Quantum mechanically,
these multi-particle states play a role even when the energy
involved in the process under study is lower than $m c^2$, because
they are summed over as virtual states in perturbation theory.
Thus, the divergences of perturbation theory in QFT are directly
linked to its short distance structure which is highly non-trivial
because its description involves the infinity of multi-particle
states.

Removing these divergences has been the nightmare and the delight
of many physicists working in particle physics. It seemed hopeless
to the non-specialist to understand renormalization because it
required prior knowledge of quantum mechanics, relativity,
electrodynamics, etc. This state of affairs contributed to the
nobility of the subject: studying the ultimate constituents
of matter and being incomprehensible fit well together. However,
strangely (at least at first sight) the theoretical breakthrough
in the understanding of renormalization beyond its algorithmic
aspect came from Wilson's work on continuous phase
transitions.\cite{wilson74} The phenomena that take
place at these transitions are neither quantum
mechanical\cite{foot1} nor relativistic and are nontrivial because
of their cooperative behavior, that is, their properties at large
distances.\cite{foot2} Thus neither
$\hbar$ nor $c$ are necessary for renormalization. Something
else is at work that does not require quantum mechanics,
relativity, summation over virtual states, Compton wavelengths,
etc., even if in the context of particle physics they are the
ingredients that make renormalization necessary. In fact, even
divergences that seemed to be the major problem of QFT are now
considered only as by-products of the way we have interpreted
quantum field theories. We know now that the invisible hand that
creates divergences in some theories is actually the existence in
these theories of a no man's land in the energy (or length) scales
for which cooperative phenomena can take place, more precisely,
for which fluctuations can add
up coherently.\cite{lepage89} In some cases,
they can destabilize the physical picture we were relying on and
this manifests itself as divergences. Renormalization, and even
more renormalization group, is the right way to deal with these
fluctuations.

One of the aims of this article is to disentangle what is specific
to field theory and what is intrinsic to the renormalization
process. Therefore, we shall not look for a physical model that 
shows
divergences,\cite{kraus92,mitra98,gosdzinsky91,mead91,adhikari97}
but we shall rather show the general mechanism of perturbative
renormalization and the renormalization group without specifying
a physical model.

\section{A toy model for renormalization}

In the following, we consider an unspecified theory that
involves, by hypothesis, only one free parameter $g_0$
in terms of which a function
$F(x)$, representing a physical quantity, is calculated
perturbatively, that is, as a power series. An example in QFT
would be quantum electrodynamics (QED), which describes the
interaction of charged particles such as electrons with the
electromagnetic field. For high energy processes, the mass of the
electron is negligible and the only parameter of this theory in
this energy regime is its charge, which is therefore the analogue
of $g_0$.
$F$ can then represent the cross section of a scattering process
as, for instance, the scattering of an electron on a heavy nucleus
in which case $x$ is the energy-momentum four-vector of the
electron. The coupling constant $g_0$ is defined by the
Hamiltonian of the system, and $F$ is calculated perturbatively
using the usual (\`a la Feynman) approach. Another important
example is continuous phase transitions. For fluids,
$F$ could represent a density-density correlation function and for
magnetism a spin-spin correlation function.\cite{foot3} Yet
another example is the solution of a differential equation that
can arise in some physical context and that can show divergences
(see the following).

It is convenient for what follows to assume that $F(x)$ has the
form:
\begin{equation}
F(x)= g_0 +g_0^2 F_1(x) + g_0^3 F_2(x) + \dots
\label{developpement}
\end{equation}
Up to a redefinition of $F$, this form is general and corresponds 
to what is really encountered in field
theory. Let us now assume that the perturbation expansion of $F(x)$ is
ill-defined and that the
$F_i(x)$ are functions involving divergent quantities. An example of such a function is:
\begin{equation}
F_1(x)= \alpha \! \int_0^\infty {dt\over t+x},
\label{exemple}
\end{equation}
which is logarithmically divergent at the upper limit. This example
has been chosen because it shares many common features with
divergent integrals encountered in QFT: the integral corresponds to
the summation over virtual states and $\alpha (t+x)^{-1}$
represents the probability amplitude associated with each of these
states.\cite{foot4}

A simple although crucial observation is that because there is only
one free parameter in the theory by
hypothesis, only one ``measurement'' of
$F(x)$, say at the point $x=\mu$, is necessary to fully specify the
theory we are studying. Such a measurement is used to fix the value of $g_0$ 
so as to reproduce the experimental value of $F(\mu)$. For QED for instance, 
this procedure would mean that:

\begin{enumerate}[(i)]

\item We start by writing
a general Hamiltonian compatible with basic 
assumptions, for example, relativity, causality, locality, and gauge
invariance.

\item We calculate physical processes at a given order of
perturbation theory, 

\item We fix the free parameter(s) of the initial Hamiltonian to
reproduce at this order the experimental data.

\end{enumerate}

This last step requires as much data as there are free parameters.
Once the parameters are fixed, the theory is completely determined
and thus predictive. One could then think that it does not matter
whether we parametrize the theory in terms of $g_0$, which
is only useful in intermediate calculations, or with a
``physical,'' that is, a measured quantity
$F(\mu)$, because $g_0$ will be replaced by this quantity anyway.
Having this freedom is indeed the generic situation in physics,
but the subtlety here is that the perturbation expansion of $F(x)$
is singular, and, thus, so is the relationship between $g_0$ and
$F(\mu)$. Thus, it seems crucial to reparametrize $F$ in terms of
$F(\mu)$ when the expansion is ill-defined.

The {\em renormalizability hypothesis} is 
that the reparametrization of the theory in terms of a physical
quantity, instead of $g_0$, is enough to turn the perturbation
expansion into a well-defined expansion. The hypothesis is
therefore that the problem does not come from the perturbation
expansion itself, that is, from the functions $F_i(x)$, but from
the choice of parameter used to perform it. This hypothesis 
means that the physical quantity, $F(x)$, initially represented by
its ill-defined expansion Eq.~(\ref{developpement}), should have a
well-defined perturbation expansion once it is calculated in terms
of the physical parameter
$F(\mu)$. This is the simplest hypothesis we can make, because
it amounts to preserving the $x$-dependence of the functions
$F_i(x)$ and only modifying the coupling constant $g_0$.
Thus, we assume that $F(x)$ is known at one point $\mu$, and we
define $g_R$ by:
\begin{equation}
F(\mu)=g_R.
\label{prescription}
\end{equation}
In the following, and by analogy
with QFT, we call $g_R$ the renormalized coupling constant
and Eq.~(\ref{prescription}) a ``renormalization prescription,'' a
barbarian name for such a trivial operation. 

We are now in a position to discuss the renormalization program. It
consists of reparametrizing the perturbation expansion of
$F$ so that it obeys the prescription of
Eq.~(\ref{prescription}). The point here is that we cannot use
Eq.~(\ref{prescription}) together with Eq.~(\ref{developpement})
because Eq.~(\ref{developpement}) is ill-defined. We first need to
give a well-defined meaning to the perturbation expansion. This is
the {\em regularization} procedure which is the first step of any
renormalization.\cite{delbourgo76,nyeo00} The idea is to define the
perturbation expansion of $F$ by a limit such that (i) the
$F_i(x)$ are well-defined before the limit is taken, and (ii)
after the renormalization has been performed, the original formal
expansion is recovered when the limit is taken. 

We thus introduce a new set of (regularized) functions $F_\Lambda$
and 
$F_{i,\Lambda}$, involving a new parameter
$\Lambda$, which we call the regulator, and such that for $\Lambda$
finite all these functions are finite. We thus define:
\begin{equation}
F_\Lambda(x)=F_\Lambda(x,g_0,\Lambda)= g_0 +g_0^2 
F_{1,\Lambda}(x) + g_0^3 F_{2,\Lambda}(x) + \dots
\label{perturbationreg}
\end{equation}
There are infinitely many ways of regularizing the $F_i$'s and for
the example given in Eq.~(\ref{exemple}), it can consist for
instance in introducing a cut-off in the integral:
\begin{equation}
F_{1,\Lambda}(x)= \alpha \int_0^\Lambda {dt\over t+x}.
\label{exempleregularise}
\end{equation}

Different regularization schemes can lead to very different
intermediate calculations, but must all lead to identical
results.\cite{foot5} For instance, dimensional regularization is
widely used in QFT because it preserves Lorentz and gauge
symmetries.\cite{hans83,gosdzinsky91,mitra98,kraus92} We do not
need here to specify a regularization for the function $F$, because
our arguments will be general and the few calculations elementary.

Once a regularization scheme has been chosen, it is possible to use
the renormalization prescription, Eq.~(\ref{prescription}),
together with the regularized expansion,
Eq.~(\ref{perturbationreg}), to obtain a well defined perturbation
series for
$F_\Lambda$ in terms of the physical coupling $g_R$. If this
expansion makes sense --- this is the renormalizability hypothesis
--- it must be finite even in the limit $\Lambda\to \infty$, because
it expresses a finite physical quantity $F(x)$ in terms of a
physical quantity
$g_R$. Thus, the renormalization program consists first in changing 
$F(x,g_0)$ to $F_\Lambda(x,g_0,\Lambda)$, then in rewriting
$F_\Lambda$ in terms of $g_R$ and $\mu$,
\begin{equation}
F_\Lambda(x,g_0,\Lambda) \to F_\Lambda(x,g_R,\mu),
\end{equation}
and only then taking the limit $\Lambda\to\infty$ at fixed $g_R$
and $\mu$. If this limit exists, $F_{\infty}(x)$ is by
hypothesis the function
$F(x)$:
\begin{equation}
F(x)=F(x,g_R,\mu)\underset{\Lambda\to\infty}{=}F_\Lambda(x,g_R,\mu).
\end{equation} 
Of course, the divergences must still be somewhere, and we
shall see that they survive in the relationship between $g_0$ and
$g_R$; at fixed $g_R$, $g_0$ diverges when $\Lambda\to\infty$. In
the traditional interpretation of renormalization, this divergence
is supposed to be harmless because $g_0$ is supposed to be a
non-physical quantity. We shall come back to this point later. 

The renormalization program is performed recursively, and we
now implement it order by order to see how it works and the
constraints on the perturbation expansion that it implies. Let us
emphasize that the series expansion we shall use in intermediate
calculations are highly formal because they are ill-defined in the
limit $\Lambda=\infty$. They are justified only by the result we
finally obtain: a good perturbation expansion in terms of
$g_R$.\cite{foot6}

$\bullet$ {\it Renormalization at order $g_0$.} At this order $F(x)$ is constant and given by:
\begin{equation}
F_\Lambda(x)= g_0 + O(g_0^2).
\end{equation} 
Thus the use of Eq.~(\ref{prescription}) leads to:
\begin{equation}
g_0=g_R+O(g_R^2).
\end{equation}

$\bullet$ {\it Renormalization at order $g_0^2$.} Our only freedom to eliminate the divergence of $F_\Lambda(x)$ is to redefine $g_0$. Because
we are working perturbatively, we expand $g_0$ as a power series
in $g_R$. Thus, we set:
\begin{equation}
g_0=g_R + \delta_2 g
+ \delta_3 g + \dots,
\end{equation}
where $\delta_n g \sim O(g_R^{n})$. At
order $g_R^2$ we obtain:
\begin{equation}
F_\Lambda(x)= g_R+ \delta_2 g +g_R^2 
F_{1,\Lambda}(x) + O(g_R^3),
\label{Fpremierordre}
\end{equation}
where we have used $g_0^2=g_R^2 +O(g_R^3)$. If we impose
Eq.~(\ref{prescription}) at this order, we obtain
\begin{equation}
\delta_2 g= - g_R^2 
F_{1,\Lambda}(\mu),
\label{delta1}
\end{equation}
which diverges when $\Lambda\to \infty$. In our example,
Eq.~(\ref{exempleregularise}), we find:
\begin{equation}
\delta_2 g= - \alpha g_R^2
\! \int_0^\Lambda {dt\over t+\mu}= -\alpha g_R^2 \log{\Lambda
+\mu\over\mu}.
\end{equation}
If we substitute Eq.~(\ref{delta1}) into Eq.~(\ref{Fpremierordre}),
we obtain $F_\Lambda$ to this order:
\begin{equation} 
F_\Lambda(x)= g_R + g_R^2 \big(F_{1,\Lambda}(x) -
F_{1,\Lambda}(\mu)\big) + O(g_R^3).
\label{Frenormalise2}
\end{equation}
It is clear that this expression for $F_\Lambda(x)$ is finite for
all
$x$ at this order if and only if the ``divergent'' part of
$F_{1,\Lambda}(x)$ (the part that becomes divergent when
$\Lambda\to\infty$) is exactly cancelled by that of
$F_{1,\Lambda}(\mu)$, that is, if and only if
\begin{equation}
F_{1,\Lambda}(x)-F_{1,\Lambda}(\mu)\ \mbox{is regular in $x$ and
$\mu$ for
$\Lambda\to\infty$.}
\label{conditionpremierordre}
\end{equation}
This condition of course means
that the divergent part of $F_{1,\Lambda}(x)$ must be a constant,
that is, is $x$-independent. If this is so, then we define the
function $F(x)$ --- now called renormalized --- as the limit of
$F_\Lambda(x)$ when
$\Lambda\to\infty$. The condition (\ref{conditionpremierordre}) is 
fulfilled for the
example of Eq.~(\ref{exemple}), and we trivially find that $F(x)$
reads:
\begin{equation}
F(x)= g_R +\alpha (\mu-x) g_R^2 \! \int_0^\infty {dt\over
(t+x)(t+\mu)} +O(g_R^3),
\end{equation}
which is obviously well defined and such that the prescription of
Eq.~(\ref{prescription}) is verified. We say that we have
renormalized the theory to this order.

Before going to the next order of perturbation theory, let us note
two important facts. First, the renormalization procedure consists
of ``adding a divergent term'' $\delta_2 g$ to $F_\Lambda$ to
remove its divergence. The cancellation takes place between the
second term of its expansion and the first one of order $g_0$.
Both lead to a term of order $g_R^2$, the one coming from the
expansion of $g_0$ in terms of $g_R$ being tuned so as to cancel
the divergence of the other. This mechanism of cancellation is a
general phenomenon: a divergence coming from the
$n$th term of the perturbation expansion is cancelled by the
expansion in powers of $g_R$ of the
$n-1$ preceding terms. Second, this cancellation is 
possible for all $x$ only if the divergence of
$F_{1,\Lambda}(x)$ is a number, that is, is
$x$-independent. If it is not so, then $F_{1,\Lambda}(x) -
F_{1,\Lambda}(\mu)$ would still be divergent $\forall x\ne\mu$.
This divergence would require the imposition of at least one
more renormalization prescription to be removed
and this second prescription would define a second, independent,
coupling constant (see Appendix~A for two functions, one 
renormalizable and one that is not). The necessity for a second
measurement of $F(x)$ would 
contradict our assumption that there is only one free parameter in
the theory. Thus we conclude that this assumption drastically
constrains the $x$-dependence of the divergences at order $g_0^2$.
We actually show in the following that this constraint propagates
to any order of perturbation theory in a non-trivial way. We also
will show that together with dimensional analysis and for a very
wide and important class of theories, these constraints are
sufficient to determine the analytical form of the divergences.

$\bullet$ {\it Renormalization at order $g_0^3$.} We suppose that
$F$ can be renormalized at order $g_R^2$, that is, condition
(\ref{conditionpremierordre}) is fulfilled. To understand the
structure of the renormalization procedure, it is necessary to go
one step further. At order $g_R^3$ we obtain:
\be
\begin{array}{ll}
F_\Lambda(x)= g_R + & \hspace{-0.1cm}{\delta{\hskip-0.3mm}}_2{\hskip-0.2mm}g\phantom{\frac{\int}{\int}}\hspace{-0.1cm} \hspace{-0.1cm} +{\delta{\hskip-0.3mm}}_3{\hskip-0.2mm}g + 
(g_R^2 + 2 g_R\, {\delta{\hskip-0.3mm}}_2{\hskip-0.2mm}g) F_{1,\Lambda}(x)+ \\
&+ g_R^3\, F_{2,\Lambda}(x) +O(g_R^4)\phantom{\int}
\end{array}
\label{Fregularise3}
\ee
where we have used $g_0^3= g_R^3 + O(g_R^4)$ and $g_0^2=g_R^2 + 2
g_R \delta_2 g+ O(g_R^4)$. We again
impose the prescription Eq.~(\ref{prescription}) and obtain:
\begin{equation}
\delta_3 g = 2 g_R^3 \big(F_{1,\Lambda}(\mu)\big)^2 - g_R^3 
F_{2,\Lambda}(\mu).
\label{this expression}
\end{equation}
If we substitute Eq.~(\ref{this expression}) in
Eq.~(\ref{Fregularise3}), we obtain:
\be
\begin{array}{l}
F_\Lambda(x)= g_R + g_R^2 \big(
F_{1,\Lambda}(x) - F_{1,\Lambda}(\mu)\big) + g_R^3 \Big( F_{2,\Lambda}(x)\\
 - F_{2,\Lambda}(\mu) - 2 F_{1,\Lambda}(\mu)\big(F_{1,\Lambda}(x) -
F_{1,\Lambda}(\mu)\big)\Big) + O(g_R^4).
\end{array}
\label{Frenormalise3}
\ee
Once again, we require that the divergence has been subtracted
for all $x$ which imposes on the
$x$-dependence of the divergent part of $F_{2,\Lambda}(x)$:
\be
\begin{array}{l}
F_{2,\Lambda}(x) - F_{2,\Lambda}(\mu) - 2 F_{1,\Lambda}(\mu)\big(F_{1,\Lambda}(x) -
F_{1,\Lambda}(\mu)\big)\ \ \  \mbox{is}\\
\ \ \ \ \ \  \mbox{ regular in $x$ and $\mu$ when $\Lambda\to\infty$}\ \ .
\end{array}
\label{condition2}
\ee
Notice that this constraint does not only involve $F_{2,\Lambda}$
but also  $F_{1,\Lambda}$. It is convenient to
rewrite
$F_{1,\Lambda}(x)$ and
$F_{2,\Lambda}(x)$ as the sum of a regular and of singular (when
$\Lambda\to\infty$) part:
\begin{equation}
F_{i,\Lambda}(x)=F_{i,\Lambda}^{s}(x)+F_{i,\Lambda}^{r}(x).
\end{equation}
Because $\infty + \mbox{anything finite} = \infty$,
this decomposition is not unique: the $F_{i,\Lambda}^{s}(x)$ are
defined up to a regular part. It is convenient to choose
$F_{1,\Lambda}^s(x)$ such that:
\begin{equation}
F_{1,\Lambda}^{s}(x)-F_{1,\Lambda}^{s}(\mu)
\underset{\Lambda\to\infty}{\longrightarrow} 0,
\label{condition1}
\end{equation}
which, of course, implies condition (\ref{conditionpremierordre}).
We show in Appendix~B that, reciprocally, this
choice is always possible if (\ref{conditionpremierordre}) is
fulfilled.  As already stated,
Eq.~(\ref{condition1}) means that the divergent part of
$F_{1,\Lambda}$ is $x$-independent. We can actually impose
a more stringent condition on $F_{1,\Lambda}^{s}$ because, by again
tuning the regular part of $F_{1,\Lambda}$, we can choose
$F_{1,\Lambda}^s$ to be completely independent of $x$, for any
$\Lambda$. We thus define:
\begin{equation}
F_{1,\Lambda}^s(x)=f_1(\Lambda).
\label{solutionf1}
\end{equation}
In our example, Eq.~(\ref{exempleregularise}), we can choose:
\begin{equation}
f_1(\Lambda)=\alpha \log \Lambda\quad \mbox{and}\quad
F_{1,\Lambda}^r(x)= \alpha \log (\frac{\Lambda+x}{\Lambda
x}).
\label{solfisfir}
\end{equation}
We now substitute Eq.~(\ref{solutionf1}) into
Eq.~(\ref{condition2}) and, using the same kind of arguments as in
Appendix~B, we obtain a constraint on the singular part of
$F_{2,\Lambda}(x)$ similar to the one on $F_{1,\Lambda}^{s}(x)$,
Eq.~(\ref{condition1}):
\begin{equation}
F_{2,\Lambda}^s(x) -F_{2,\Lambda}^s(\mu) - 2 f_1(\Lambda)
\big[F_{1,\Lambda}^r(x)
-F_{1,\Lambda}^r(\mu)\big]\underset{\Lambda\to\infty}
{\longrightarrow} 0.
\label{contrainte2prime}
\end{equation}
Equation~(\ref{contrainte2prime}) can be rewritten as:
\begin{equation}
[F_{2,\Lambda}^s(x)- 2 f_1(\Lambda) F_{1,\Lambda}^r(x)]
- [F_{2,\Lambda}^s(\mu)- 2 f_1(\Lambda)
F_{1,\Lambda}^r(\mu)]
\underset{\Lambda\to\infty}{\longrightarrow} 0.
\label{same structure}
\end{equation}
Equation~(\ref{same structure}) has the same structure as
Eq.~(\ref{condition1}) up to the replacement: $F_{1,\Lambda}^s\to
F_{2,\Lambda}^s- 2 f_1(\Lambda) F_{1,\Lambda}^r$ and therefore
has the same kind of solution as Eq.~(\ref{solutionf1}):
\begin{equation}
F_{2,\Lambda}^s(x) = 2 f_1(\Lambda) F_{1,\Lambda}^r(x) +
f_2(\Lambda),
\label{last}
\end{equation}
where $f_2(\Lambda)$ is any function of $\Lambda$ and is
independent of $x$. We see in Eq.~(\ref{last}) that unlike
$F_{1,\Lambda}(x)$, the divergent part of $F_{2,\Lambda}(x)$
depends on $x$. However, this dependence is entirely determined by
the first order of the perturbation expansion. The 
$\delta_2g$ term, necessary to remove the $O(g_0^2)$ divergence, has 
produced at order $g_R^3$ an $x$-dependent divergent term: $2
g_R \delta_2 g F_{1,\Lambda}(x)$.
This kind of $x$-dependence is also a general phenomenon of
renormalization: the (counter-)terms that remove divergences at a
given order produce divergences at higher orders. If the theory is
renormalizable, these divergences contribute to the cancellation
of divergences present in the perturbation expansion at higher
orders. Thus, perturbative renormalizability, that is, the
possibility of eliminating order by order all divergences by the redefinition
of the coupling(s),
implies a precise structure of (the divergent parts of) the
successive terms of the perturbation series. At order $n$, the
singular part of
$F_{n,\Lambda}$ involves $x$-dependent terms entirely determined
by the preceding orders plus one new term that is
$x$-independent. In our example of Eq.~(\ref{exemple}) and
Eq.~(\ref{solfisfir}) we find:
\begin{equation}
F_{2,\Lambda}^s(x) = 2 \alpha^2 \log\Lambda
\log\frac{x+\Lambda}{\Lambda x}+ f_2(\Lambda).
\end{equation}
By expanding $\log\Lambda \log(x+\Lambda)/{\Lambda x}$ in powers
of $\Lambda^{-1}$ and by again redefining the regular part of
$F_{2,\Lambda}$, we obtain a simpler form for $F_{2,\Lambda}^s(x)$:
\begin{equation}
F_{2,\Lambda}^s(x) = -2 \alpha^2 \log\Lambda \log x+
f_2(\Lambda).
\label{solution2exemple}
\end{equation}
This relation will be important in the following when we shall
discuss the renormalization group. 

Let us draw our first conclusion. Infinities occur in the
perturbation expansion of the theory because we have assumed that
it was not regularized. Actually, these divergences have forced us
to regularize the expansion and thus to introduce a new scale
$\Lambda$. Once regularization has been performed, renormalization
can be achieved by eliminating $g_0$.
The limit $\Lambda\to\infty$ can then be taken. The process is
recursive and can be performed only if the divergences possess,
order by order, a very precise structure. This structure
ultimately expresses that there is only one coupling constant to
be renormalized. This means that imposing only one prescription at 
$x=\mu$ is enough to subtract the divergences for all $x$. In general, a
theory is said to be renormalizable if all divergences can be
recursively subtracted by imposing as many prescriptions as there
are independent parameters in the theory. In QFT, these are 
masses, coupling constants, and the normalization of the fields. An
important and non-trivial topic is thus to know which parameters
are independent, because symmetries of the theory (like gauge
symmetries) can relate different parameters (and Green functions). 

Let us once again recall that renormalization is nothing but a
reparametrization in terms of the physical quantity
$g_R$.\cite{foot7} The price to pay for renormalizing $F$ is that
$g_0$ becomes infinite in the limit
$\Lambda\to\infty$, see Eq.~(\ref{delta1}). We again emphasize
that if $g_0$ is believed to be no more than a non-measurable
parameter, useful only in intermediate calculations, it is indeed
of no consequence that this quantity is infinite in the limit
$\Lambda\to\infty$. That $g_0$ was a divergent non-physical
quantity has been common belief for decades in QFT. The physical
results given by the renormalized quantities were thought to be
calculable only in terms of unphysical quantities like $g_0$
(called bare quantities) that the renormalization algorithm 
could only eliminate afterward. It was as if we had to make two
mistakes that compensated each other: first introduce bare
quantities in terms of which everything was infinite, and then
eliminate them by adding other divergent quantities. Undoubtly,
the procedure worked, but, to say the least, the interpretation
seemed rather obscure.

Before studying the renormalization group, let us now specialize to
a particular class of renormalizable theories.

\section{Renormalizable theories with dimensionless couplings}
A very important class of field theories corresponds to the
situation where $g_0$ is dimensionless, and $x$, which in QFT
represents coordinates or momenta, has dimensions (or more
generally when $g_0$ and $x$ have independent dimensions). In
four-dimensional space-time, quantum
electrodynamics is in this class, because the fine structure
constant is dimensionless; quantum chromodynamics
and the Weinberg-Salam model of electro-weak interactions are
also in this class. In four space
dimensions, the $\phi^4$ model 
relevant for the Ginzburg-Landau-Wilson approach to critical
phenomena is in this class too. This particular class of renormalizable
theories is the cornerstone of renormalization in field theories.

Our main goal in this section is to show that, independently of
the underlying physical model, dimensional analysis together with
the renormalizability constraint, determine almost entirely the
structure of the divergences. This underlying simplicity of the
nature of the divergences explains that there is no
combinatorial miracle of Feynman diagrams in QFT as it might seem
at first glance. Let us now see in detail how it works.

Because $F_\Lambda(x)$ has the same dimension as $g_0$, it
also is dimensionless and so are the
$F_{i,\Lambda}(x)$. The only possibility for a dimensionless
quantity like $F$ to be a function of a dimensional variable like
$x$ is that there exists another dimensional variable such that
$F$ depends on $x$ only through the ratio of these two variables.
Apart from $x$, the only other quantity on which $F$ depends is
$\Lambda$, which must therefore have the same dimension as $x$.
This is indeed the case in our example,
Eq.~(\ref{exempleregularise}). Thus, the functions
$F_{i,\Lambda}(x)$ depend on the ratio $x/\Lambda$
only.\cite{foot8} Let us show that this is enough to
prove that the $F_{i,\Lambda}^s(x)$ are sums of powers of
logarithms with, for most of them, prescribed prefactors.

Let us start with $F_{1,\Lambda}^s(x)$. On one hand, we have seen
that by redefining the regular part of $F_{1,\Lambda}(x)$, we
could take its singular part $F_{1,\Lambda}^s(x)$ independent of
$x$, Eq.~(\ref{solutionf1}). On the other hand, we know that 
$F_{1,\Lambda}(x)$ is a function of $x/\Lambda$. Thus, by
redefining $F_{1,\Lambda}^r(x)$, it must be possible to extract an
$x$-dependent regular part, $r(x)$, of this function so as to build the 
$x/\Lambda$ dependence of $F_{1,\Lambda}^s(x)$:
\begin{equation}
F_{1,\Lambda}^s(x)=f\big(\frac{x}{\Lambda}\big)=f_1(\Lambda) +
r(x).
\label{eq29}
\end{equation}
Hence, $F_{1,\Lambda}^s$ is separable into functions of $x$ only and of $\Lambda$ only
which sum up to a function of $x/\Lambda$. We show in Appendix~C
the well known fact that only the logarithm obeys this property.
We obtain (see Eqs.~(\ref{c3}) and (\ref{eq2}))
\begin{equation}
F_{1,\Lambda}^s(x)=-f_1\big(\frac{x}{\Lambda}\big)=f_1(\Lambda)-f_1(x)=
\alpha
\log\frac{\Lambda}{x}.
\label{eq30}
\end{equation}
Therefore, for renormalizable theories and for dimensionless
functions such as $F$, only logarithmic divergences are allowed at
order
$g_0^2$ (in QFT, this is the so-called one-loop term). This is the
reason why logarithms are encountered everywhere in QFT. Note
that because of dimensional analysis, the finite part of
$F_{1,\Lambda}(x)$ is nothing but
$r(x)$, up to an additive constant, at least for
$\Lambda\to\infty$. This can be checked for the example given in
Eq.~(\ref{exempleregularise}). Thus, by dimensional analysis, the
structure of the divergence determines that of the finite part (up
to a constant). Notice that things would not be that simple if
$F_{\Lambda}(x)$ depended on another dimensional parameter, which
is the case of massive field theories where masses and momenta 
have the same dimension. In this case, the finite part is only
partially determined by the singular one.

Let us now show that the structure of $F_{2,\Lambda}^s$ also is 
entirely determined for renormalizable theories with
dimensionless couplings both by the renormalizability hypothesis
and by dimensional analysis. We have already partially studied
this case with the example given in Eq.~(\ref{exempleregularise})
where $F_{1,\Lambda}^s(x)$ is logarithmically divergent, a
characteristic feature of these renormalizable theories. In
particular, we have shown that in this case, renormalizability
imposes at order $g_0^3$ that $F_{2,\Lambda}^s$ is of the form
given in Eq.~(\ref{solution2exemple}). Let us now use dimensional
analysis that once again imposes that $F_{2,\Lambda}^s$ depends
only on $x/\Lambda$. The only freedom we have to reconstruct a
function of $x/\Lambda$ from the form of $F_{2,\Lambda}^s$ given
in Eq.~(\ref{solution2exemple}) is to add a regular function to
it. It is not difficult to find how to proceed because the only
admissible term including $\log\Lambda \log x$ is
$\log^2\Lambda/x$:
\begin{equation}
\log^2\frac{\Lambda}{x}= \log^2 \Lambda - 2 \log\Lambda \log x +
\log^2 x.
\end{equation}
Thus, to obtain the dimensionally correct extension of the term $-2
\alpha^2 \log\Lambda \log x$ in Eq.~(\ref{solution2exemple}), we 
extract $\alpha^2\log^2 \Lambda$ from $f_2(\Lambda)$ and add the
regular term $\alpha^2\log ^2x$:
\be
\begin{array}{ll}
-2& \alpha^2 \log\Lambda \log x + f_2(\Lambda)\\
&\displaystyle{\rightarrow\  -2 \alpha^2 \log\Lambda \log x + \alpha^2 \log^2\Lambda +\big(f_2(\Lambda)-\alpha^2 \log^2\Lambda \big)\phantom{\int}}\\
&\rightarrow\  -2 \alpha^2 \log\Lambda \log x + \alpha^2 \log^2\Lambda + \alpha^2 \log^2 x\, + \\
&\ \ \ \ \ \ \ \ \ \ +\big(f_2(\Lambda)-\alpha^2 \log^2\Lambda \big)\phantom{\int}\\
&\displaystyle{\rightarrow\  \alpha^2\log^2\frac{\Lambda}{x}+ \big(f_2(\Lambda)-\alpha^2 \log^2\Lambda \big)\phantom{\frac{\int}{\int}}}\ \ .
\end{array}
\ee
Thus, we obtain for the new function $F_{2,\Lambda}^s(x)$:
\begin{equation}
F_{2,\Lambda}^s(x)= \alpha^2 \log^2\frac{\Lambda}{x} +
f_2(\Lambda)-\alpha^2
\log^2\Lambda.
\end{equation}
Now, for $f_2(\Lambda)-\alpha^2 \log^2\Lambda$, we can repeat the
same argument as the one used previously for 
$F_{1,\Lambda}^s(x)$ (which is equal to 
$f_1(\Lambda)$, Eq.~(\ref{solutionf1})): it is a function of
$\Lambda$ that must become a function of $x/\Lambda$ only by
adding a function of $x$. It is thus also a logarithm, see
Eqs.~(\ref{eq29}) and (\ref{eq30}) and Appendix~C. Therefore,
we add a $\log x$ term to
$F_{2,\Lambda}^s(x)$ and obtain the final result:
\begin{equation}
F_{2,\Lambda}^s(x)= \alpha^2 \log^2\frac{\Lambda}{x} + \beta
\log\frac{\Lambda}{x},
\label{solutionf2exemple}
\end{equation}
where $\beta$ is a pure number. We emphasize that although it is
$x$-independent, the term $\alpha^2 \log^2\Lambda$ involved in
$F_{2,\Lambda}^s(x)$ arises from the $\log\Lambda \log x$ term
thanks to dimensional analysis. It is thus entirely determined by
the term of order $g_0^2$ of perturbation theory. Only the
sub-leading logarithm $\beta \log\Lambda/x$ is new. It is not
difficult now to guess the structure of the next order of
perturbation: it involves a $\log^3{\Lambda}/{x}$ with a prefactor
$\alpha^3$, a $\log^2{\Lambda}/{x}$ term with a prefactor which
is a function of $\alpha$ and $\beta$ and a $\log{\Lambda}/{x}$
with a prefactor independent of $\alpha$ and $\beta$. A precise
calculation shows that:
\begin{equation}
\begin{array}{lll}
F_{\Lambda}^s(x)={\displaystyle\alpha g_0^2\log\frac{\Lambda}{x}} 
       &{\displaystyle+ \alpha^2 g_0^3 \log^2\frac{\Lambda}{x}} &{\displaystyle+\alpha^3g_0^4 \log^3\frac{\Lambda}{x}+\dots} \\
       &{\displaystyle+  \beta g_0^3 \log\frac{\Lambda}{x}}     &{\displaystyle+\frac{5}{2}\alpha\beta g_0^4\log^2\frac{\Lambda}{x}+\dots}\\
       &                                                        &{\displaystyle+\gamma g_0^4\log\frac{\Lambda}{x}+\dots}
\end{array}
\label{serief}
\end{equation}
We have written the series so as to exhibit its ``triangular''
nature: the first line corresponds to the leading logarithms, the
second to the sub-leading, etc., and the $n$th column to the 
$n$th order of perturbation. The leading logarithms are
entirely controlled by the $g_0^2$ term, the sub-leading
logarithms by both the
$g_0^2$ and $g_0^3$ terms, etc. It is clear that order by order
for the divergent terms, only the $\log$ term is new, all the
$\log^2$, $\log^3$, etc, terms are determined by the preceding
orders. This structure strongly suggests that we can,
at least partially, resum the perturbation series. We notice
that although the leading logarithms form a simple geometric
series, this is no longer true for the sub-leading logarithms
where, for instance, the factor 5$\alpha\beta/2$ of
Eq.~(\ref{serief}) is non-trivial. Thanks to the renormalization
group, there exists a systematic way to perform these
resummations\cite{shirkov01} (see the following).

We again emphasize that for our simple toy model the divergences
together with dimensional analysis determine almost entirely the
entire function $F(x)$ in the limit of large
$\Lambda$. To show this explicitly, we rewrite
$F$ as:
\begin{equation}
F_\Lambda(x,g_0,\Lambda)= g_0 + F_\Lambda^s(x,g_0,\Lambda)+ F_\Lambda^r(x,g_0,\Lambda)
\label{decomposition}
\end{equation}
with $F_\Lambda^s(x,g_0,\Lambda)$ given by Eq.~(\ref{serief}) at
$O(g_0^4)$ and $F_\Lambda^r(x,g_0,\Lambda)\sim O(g_0^2)$. From
dimensional analysis, $F_\Lambda^r(x,g_0,\Lambda)$ is also a
function of $x/\Lambda$ only which, by definition, is finite when
$\Lambda\to\infty$. Thus, for large $\Lambda$:
\begin{equation}
F_\Lambda^r(x,g_0,\Lambda)= {\cal
F}\big(\frac{x}{\Lambda},g_0\big)\simeq {\cal F}(0,g_0).
\end{equation}
$F_\Lambda^r(x)$ is therefore almost $x$-independent for large
$\Lambda$: it is a ($g_0$-dependent) number in this limit. For the
sake of simplicity, let us consider the case where it is vanishing:
\begin{equation}
F_\Lambda(x,g_0,\Lambda)= g_0 + F_\Lambda^s(x,g_0,\Lambda)
\label{fraccourci}
\end{equation}
with $F_\Lambda^s(x,g_0,\Lambda)$ a function of $x/\Lambda$ only.
By using the renormalization prescription
Eq.~(\ref{prescription}), we can calculate $g_R$ as a function of
$g_0$ and $\Lambda/\mu$ and by formally inverting the series, we
obtain at $O(g_R^4)$:
\begin{eqnarray}
g_0&=& g_R -\alpha g_R^2 \log \frac{\Lambda}{\mu}
+g_r^3 [\alpha^2 \log^2 \frac{\Lambda}{\mu} - 
\beta \log \frac{\Lambda}{\mu}] \nonumber \\
&&{} + g_R^4[-\gamma \log \frac{\Lambda}{\mu} +\frac{5}{2}
\alpha\beta \log^2 \frac{\Lambda}{\mu} -\alpha^3 \log^3
\frac{\Lambda}{\mu}].
\end{eqnarray}
By substituting this expression in Eqs.~(\ref{serief}) and
(\ref{fraccourci}), we obtain at $O(g_R^4)$:
\begin{equation}
\begin{array}{lll}
F_{\Lambda}(x)=g_R+{\displaystyle\alpha g_R^2\log\frac{\mu}{x}} 
       &{\displaystyle+ \alpha^2 g_R^3 \log^2\frac{\mu}{x}} &{\displaystyle+\alpha^3g_R^4 \log^3\frac{\mu}{x}} \\
       &{\displaystyle+  \beta g_R^3 \log\frac{\mu}{x}}     &{\displaystyle+\frac{5}{2}\alpha\beta g_R^4\log^2\frac{\mu}{x}}\\
       &                                                        &{\displaystyle+\gamma g_R^4\log\frac{\mu}{x}}
\end{array}
\label{Ftot_ren}
\end{equation}
Thus, we find that the renormalization process leaves unchanged
the functional form of $F_{\Lambda}$, Eq.~(\ref{serief}), and just consists in replacing
$(g_0,\Lambda)$ by $(g_R,\mu)$. This very important fact is
related to a self-similarity property that we study in detail from
the renormalization group viewpoint. Notice that of course any
explicit dependence on $\Lambda$ and $g_0$ has been eliminated in
Eq.~(\ref{Ftot_ren}) and that the limit $\Lambda\to\infty$ can now
be safely taken, if desired.

Note that we have obtained logarithmic divergences because we have
studied the renormalization of a dimensionless coupling constant.
If $g_0$ was dimensional, we would have obtained power law
divergences. This is for instance what happens in QFT for the mass
terms (see also in the following the expansion in
Eq.~(\ref{equa_diff_premier_ordre})).

\section{Renormalization group}
Although the renormalization group will allow us to partially
resum the perturbation expansion, we shall not introduce it in
this way. Rather, we want to examine the internal consistency of
the renormalization procedure.

We have chosen a renormalization prescription at the point $x=\mu$
where $g_R$ is defined. Obviously, this point is not special, and
we could have chosen any other point
$\mu'$ or $\mu''$ to parametrize the theory. Because there is only
one independent coupling constant, the different coupling constants
$g_R=g_R(\mu)$, $g_R'=g_R(\mu')$, $g_R''=g_R(\mu'')$ should all be
related in such a way that
$F(x)=F(x,\mu,g_R)=F(x,\mu',g_R')=F(x,\mu'',g_R'')$, etc. 
This means that there should exist an equivalence class of
parametrizations of the same theory and that it should not matter
in practice which element in the class is chosen. This independence
of the physical quantity with respect to the choice of
prescription point also means that the changes of parametrizations
should be a (renormalization) group law: going from the 
parametrization given by $(\mu,g_R)$ to that given by
$(\mu',g_R')$ and then to that given by $(\mu'',g_R'')$ or going
directly from the first parametrization $(\mu,g_R)$ to the
last one
$(\mu'',g_R'')$ should make no difference.
\begin{figure}[htbp] 
\begin{center}
\includegraphics[width=.9\linewidth,origin=tl]{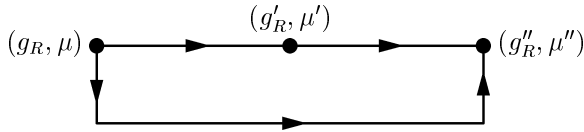}\hfill%
\end{center}
\caption{An illustration of the renormalization group: the two equivalent ways to compose changes of parametrizations.}
\label{diagRG}
\end{figure}
Put this way, this statement seems to be 
void. Actually, it is. More precisely, it would be so if
we were performing exact calculations: we would gain no new
physical information by implementing the renormalization group
law. This is because this group law does not reflect a symmetry
of the physics, but only of the parametrization of our solution.
This situation is completely analogous to what happens for
the solution of a differential equation: we can parametrize it at
time $t$ in terms of the initial conditions at time $t_0$ for
instance, or we can use the equation itself to calculate the
solution at an intermediate time $\tau$ and then use this solution
as a new initial condition to parametrize the solution at time
$t$. The changes of initial conditions that preserve the final
solution can be composed thanks to a group law. Let us consider
for example the following trivial, but illuminating, example:
\begin{equation}
\dot{y}(t)=\epsilon y(t),\ \ y(t_0)=r_0,
\label{trivial_equa_diff}
\end{equation}
the solution of which is:
\begin{equation}
y(t)=f(r_0,t-t_0)=r_0 e^{\epsilon(t-t_0)}.
\label{def_fonction_f}
\end{equation}
The group law can be written as\cite{foot9}:
\begin{equation}
f(r_0,t-t_0)= f\big(f(r_0,\tau-t_0),t-\tau\big)\ \ \ \forall \tau.
\label{loi_de_groupe}
\end{equation}
which you can verify using the exact solution,
Eq.~(\ref{def_fonction_f}). The non-trivial point with these group
laws is that, in general, they are violated at any finite order of
the perturbation expansions. In our previous example, we obtain to
order $\epsilon$:
\begin{equation}
y(t)\simeq f_1(r_0,t-t_0)=r_0 (1+\epsilon(t-t_0)),
\label{equa_diff_premier_ordre}
\end{equation}
and 
\begin{equation}
f_1\big(f_1(r_0,\tau-t_0),t-\tau\big)=
r_0 (1+\epsilon(t-t_0))+\epsilon^2 r_0
(t-\tau)(\tau-t_0).
\label{loi_groupe_approchee}
\end{equation}
The group law is verified to order $\epsilon$ because the
perturbation expansion is exact at this order. However, it is
violated by a term of order $\epsilon^2$ that can be arbitrarily
large even for small $\epsilon$, provided $t-t_0$ is large enough.

The interest of the group law, Eq.~(\ref{loi_de_groupe}), is that
it is possible to enforce it and then to improve the perturbation
result. Actually, when renormalization is necessary, the group
laws lets us partially resum the perturbation series of divergent
terms.

Let us now see how this improvement of the perturbation series works for the example of
the differential equation (\ref{trivial_equa_diff}). In this case, the divergence occurs for
$t_0\to-\infty$. Thus, $t_0$ plays the role of the cut-off
$\Lambda$, $t-t_0$ of $\log\Lambda/\mu$, and 
$t-\tau$ of $\log\mu'/\mu$. Once $t_0$ is finite, no divergence
remains, but the relics of the divergences occurring for
$t_0\to-\infty$ are the large violations of the group law because
both the divergences and these violations originate in the fact
that the perturbation expansion is performed in powers of
$\epsilon(t-t_0)$ and not of $\epsilon$. To further study the
relevance of the group law, it is interesting to forget the higher
order terms of the perturbation expansion for a while and to look
for an improved approximation that coincides at order
$\epsilon$ with the perturbation result and that obeys the
group law at order
$\epsilon^2$:
\begin{equation}
f_1^{\rm imp}(r_0,t-t_0)=r_0(1+\epsilon(t-t_0)+\epsilon^2
G(t-t_0)).
\label{fimp}
\end{equation}
By imposing the group law, Eq.~(\ref{loi_de_groupe}), to order
$\epsilon^2$, we obtain a functional equation for $G$:
\begin{equation}
G(t-t_0)=G(\tau-t_0)+ G(t-\tau) + (\tau-t_0)(t-\tau).
\label{equationgg}
\end{equation}
If we differentiate Eq.~(\ref{equationgg}) with respect to $t_0$
and take $t_0=\tau$, we obtain, setting $x=t-\tau$:
\begin{equation}
G'(x)=x+G'(0).
\label{G}
\end{equation}
Because $G(0)=0$, Eq.~(\ref{G}) implies that:
\begin{equation}
G(x)=\frac{x^2}{2}+ a x,
\label{solution_fimp}
\end{equation}
where $a$ is arbitrary. For $a=0$, this result is
actually the perturbation result to order $\epsilon^2$ because:
\begin{equation}
y(t)\simeq r_0 (1+\epsilon(t-t_0) + \frac{\epsilon^2}{2}
(t-t_0)^2) + O(\epsilon^3).
\end{equation}
Thus, the first order in the perturbation expansion, together with
the group law, determines entirely the term of highest degree in
$t-t_0$ at the next order. Of course, to verify exactly the
group law, we should pursue the expansion in $\epsilon$ to all
orders. It is easy to show that to order
$\epsilon^n$, the term of highest degree in $t-t_0$ is completely
determined by both the first order result and the group law and
coincides with the perturbation result: $\epsilon^n(t-t_0)^n/n!$.
Thus, the only information given by the perturbation expansion is
that all subdominant terms, $\epsilon^n(t-t_0)^p$ with $p<n$, 
vanish in this example. We could now show how the
implementation of the group law lets us resum the perturbation
expansion. Unfortunately, this example is too simple and some
important features of the renormalization group are missed in this
case. (See Appendix~\ref{appendix_E} for a complete discussion of
the implementation of the renormalization group on this example.)
We therefore go back to our toy model for which we specialize to
renormalizable theories with dimensionless couplings. 

\subsection{Renormalization group for renormalizable theories
with dimensionless couplings}

We now reconsider our toy model, Eqs.~(\ref{perturbationreg}),
(\ref{serief}), and (\ref{decomposition}), from the point of view
of the renormalization group. For the sake of simplicity, we keep
only the dominant terms at each order, that is, apart from $g_0$,
the divergent ones in Eq.~(\ref{fraccourci}). 

First, notice that in the same way $g_R$ is clearly associated with the
scale
$\mu$, Eq.(\ref{prescription}), so is $g_0$ with the scale $\Lambda$ because from 
Eq.~(\ref{serief}), we find\cite{foot10}:
\begin{equation}
F_\Lambda(x=\Lambda)=g_0.
\label{eqfoot10}
\end{equation}
Let us define a third coupling constant associated with the scale
$\mu'$,
\begin{equation}
F_\Lambda(\mu')=g_R',
\label{prescriptionprime}
\end{equation}
and study the relationship between these different coupling
constants at order $g_0^2$. From:
\begin{equation}
F_\Lambda(x,g_0,\Lambda)=g_0 + \alpha g_0^2
\log\big(\frac{\Lambda}{x}\big) + O(g_0^3),
\label{fordre2}
\end{equation}
we obtain:
\begin{eqnarray}
g_R&=&g_0 +\alpha g_0^2 \log\big(\frac{\Lambda}{\mu}\big) +
O(g_0^3)\label{grp.a}\\
g_R'&=&g_0 +\alpha g_0^2 \log\big(\frac{\Lambda}{\mu'}\big) +
O(g_0^3).
\label{grp}
\end{eqnarray}
By eliminating $g_0$ between these two equations, we find:
\begin{equation}
g_R'=g_R +\alpha g_R^2 \log\big(\frac{\mu}{\mu'}\big) + O(g_R^3),
\label{grgrprime}
\end{equation}
and thus, as expected, the group law controlling the change of
prescription point is verified perturbatively. We note that
the essential ingredient for this composition law is that
Eq.~(\ref{grgrprime}) is independent of $\Lambda$. This is what
ensures that the same form can be used to change
$(g_0,\Lambda)$ into $(g_R,\mu)$ and then $(g_R,\mu)$ into
$(g_R',\mu')$. This independence, in turn, is nothing but the
signature of perturbative renormalizability which lets us
completely eliminate at each order $(g_0,\Lambda)$ for $(g_R,\mu)$.
Perturbatively, everything looks fine. However, the previous
calculation relies on a formal step that is not mathematically
correct, at least for large $\Lambda$. Indeed, to go from
Eq.~(\ref{grp}) to Eq.~(\ref{grgrprime}), the series
$g_R=g_R(g_0)$ must be inverted to find $g_0=g_0(g_R)$ while, for
$\Lambda\to\infty$, the series $g_R=g_R(g_0)$ is clearly not 
convergent and thus not invertible. Thus, the neglected terms of
order $g_R^3$ in Eq.~(\ref{grgrprime}) involve a term proportional
to $\log\Lambda/\mu \log\mu'/\mu$ --- analogous to the term
$(t-\tau)(\tau-t_0)$ of Eqs.(\ref{loi_groupe_approchee}) and
(\ref{equationgg}) --- which is neglected because it is of order
$g_R^3$, but which is very large for large $\Lambda$ (see
Appendix~D). From a practical point of view, the existence at any
order of these large terms of higher orders spoil the group law 
so that the independence of the physical results with respect to
the choice of prescription point is not verified. 

As in the case of the differential equation (\ref{fimp}),
we can look for an improved function: $F^{\rm imp}$,
\begin{equation}
F^{\rm imp}(x,g_0,\Lambda)=g_0 + \alpha g_0^2
\log\big(\frac{\Lambda}{x}\big) + g_0^3
G\big(\frac{\Lambda}{x}\big)+ O(g_0^4),
\end{equation}
for which the group law at order $g_0^3$ is obeyed. It is shown
in Appendix~D that this constraint implies that:
\begin{equation}
G(x)=\alpha^2 \log^2x +\beta \log x,
\label{equation_G}
\end{equation}
where $\beta$ is arbitrary. Thus:
\be
\begin{array}{ll}
\displaystyle{F^{\small\mbox{imp}}(x,g_0,\Lambda)}=& \hspace{-0.1cm}g_0 +\displaystyle{ \alpha g_0^2 \log\left(\frac{\Lambda}{x}\right) + 
\alpha^2 g_0^3 \log^2\left(\frac{\Lambda}{x}\right)+} \\
&\ +\,\displaystyle{\beta g_0^3 \log\left(\frac{\Lambda}{x}\right) +O(g_0^4)}\ \ .
\end{array}
\ee
Once again, we find that the group law together with the order
$g_0^2$ result determines the leading behavior at the next order,
here the $\log^2 (\Lambda/x)$ term. Moreover,
we find that the group law imposes the existence of the same
log$^2$ term as the one found from the renormalizability
constraint, Eqs.(\ref{solutionf2exemple}) and (\ref{serief}), and
allows the existence of a sub-leading logarithm. Although
non-trivial, this should not be too surprising because the
renormalizability constraint means that once $F$ is well defined
at $x=\mu$, it also is everywhere and in particular at $x=\mu'$.
The renormalizability constraint is therefore certainly necessary for the
implementation of the group law. As in the example of the
differential equation, Eq.~(\ref{trivial_equa_diff}), we should
pursue the expansion to all orders to obtain an exactly verified
group law. It is clear that by doing so, we would find the same
expansion as the one obtained from the renormalizability
constraint. Thus, if we use perturbation theory to calculate the
coefficient in front of the 
first leading logarithm (of order $g_0^2$) and impose the group
law, we should be able to resum all the leading logarithms. To do
the resummation of the sub-leading and sub-sub-leading logarithms,
a knowledge of respectively the order $g_0^3$ and $g_0^4$ terms
is required. Clearly, we need to understand how
to systematically construct the function
$f$ giving
$g_R'$ in terms of $g_R$ and $\mu/\mu'$,\cite{shirkov01}
\begin{equation}
g_R'= f\big(g_R,\frac{\mu}{\mu'}\big),
\label{fonctionfgr}
\end{equation}
such that 
\begin{itemize}

\item its expansion at order $n$ is given by the $n$th order of
perturbation theory, 

\item the group law is {\it exactly} verified:
\begin{equation}
f\big(g_R,\frac{\mu}{\mu''}\big)=
f\big(f\big(g_R,\frac{\mu}{\mu'}\big),\frac{\mu'}{\mu''}\big).
\label{grouplaw}
\end{equation}
\end{itemize}
The function $f$ is then said to be the {\em self-similar
approximation} at order $n$ of the exact relationship between
$g_R$ and $g_R'$.\cite{kovalev99} First notice one crucial thing.
Our first aim was to study the perturbation expansion of a
function $F$ in a power series of a coupling constant $g_0$. Then
we have discovered that the logarithmic divergence at order
$g_0^2$ propagates to all orders so that the expansion is actually
performed in $g_0 \log\Lambda/\mu$ instead of $g_0$. Because
$\Lambda$ is the regulator, it is supposed to be very large
compared with $\mu$, so that the large logarithmic terms
invalidate the use of the perturbation expansion. Reciprocally, it
is clear that perturbation theory is perfectly valid if it is
performed between two scales $\mu_1$ and $\mu_2$ which are very
close. Thus, instead of using perturbation theory to make a big
jump between two very distinct scales, say $\Lambda$ and $\mu$, we
should use it to perform a series of very little steps for which
it is valid at each of them. In geometrical terms, the fact that
the perturbative approach is valid only between two very close scales
 means that we should not use perturbation theory to
approximate the equation of the curve given by the function $f$,
Eq.~(\ref{fonctionfgr}), that joins the points $(\mu,g_R)$ and
$(\mu',g_R')$, but we should use it to calculate the (field of)
tangent vectors to this curve, that is, its envelope. The curve
itself should then be reconstructed by integration, see
Appendix~\ref{appendix_E}. By doing so, the group law will be
automatically verified because, by construction, the integration
precisely consists in composing infinitesimal changes of
reparametrization infinitely many times. Let us consider again
Eq.~(\ref{grp.a}). We want to calculate the evolution of 
$g_R(\mu)$ with $\mu$ for a given model specified by
$(\Lambda,g_0)$. Thus we define:
\begin{equation} 
\beta(g_R)=\mu\frac{\partial
g_R}{\partial\mu}\bigg|_{g_0,\Lambda},
\label{beta}
\end{equation}
which gives the infinitesimal evolution of the coupling constants
with the scale for the model corresponding to 
$(g_0,\Lambda)$. We trivially find to this order from
Eq.~({\ref{grp.a}),
\begin{equation}
\beta(g_R) =-\alpha g_0^2 +O(g_0^3),
\label{betag0}
\end{equation}
and thus, by trivially inverting the series of Eq.~(\ref{grp.a}), we
obtain:
\begin{equation}
\beta(g_R) =-\alpha g_R^2 +O(g_R^3).
\label{betaordre1}
\end{equation}
Now, if we integrate Eq.~(\ref{beta}) together with
Eq.~(\ref{betaordre1}), we obtain
\begin{equation}
g_R'=\frac{g_R}{1-\alpha g_R \log\frac{\mu}{\mu'}}.
\label{grgrprimeintegre}
\end{equation}
This relation has several interesting properties:

\begin{enumerate}[(i)]

\item When expanded to order $g_R^2$, the perturbation result
to this order is recovered, Eq.~(\ref{grgrprime}). This is quite
normal because $\beta(g_R)$ has been calculated to this order.

\item When expanded to all orders, the whole series of leading
logarithms is recovered. This is more interesting because
$\beta(g_R)$ has been calculated only to order $g^2$, but 
simply means that all the leading logarithms are
determined by the first one.

\item The group law (\ref{grouplaw}) is obeyed exactly. We
have thus found the function $f$ of Eq.~(\ref{fonctionfgr}) to
this order. It is very instructive to check the group law directly
from Eq.~(\ref{grgrprimeintegre}) and to verify that the
$\beta$-function found in Eq.~(\ref{betaordre1}) is not
modified if we add the leading logarithmic term of order $g_0^3$
to the relation (\ref{grp.a}): 
\begin{equation}
g_R=g_0 +\alpha g_0^2 \log\big(\frac{\Lambda}{\mu}\big)+\alpha^2
g_0^3 \log^2\big(\frac{\Lambda}{\mu}\big) + O(g_0^4).
\label{relationordre3grg0}
\end{equation}
The independence of the $\beta$-function with respect to the addition of
the successive leading logarithmic terms means that this function is indeed
the right object to build self-similar approximations out of the
perturbation expansion. 

\end{enumerate}

Let us now return to the $\beta$-function itself. First, we have
calculated the logarithmic derivative
$\mu\partial g_R/\partial\mu$ instead of the ordinary derivative
with respect to $\mu$ because we wanted
to have a dimensionless $\beta$-function. Second, even the
dimensionless quantity,
$\beta(g_R)$, could have depended on $\Lambda/\mu$.
However, the evolution of $g_R(\mu)$ between $\mu$ and
$\mu+d\mu$ cannot depend in perturbation theory on
$\Lambda$ because the theory is perturbatively renormalizable: the
perturbative relation between $g_R(\mu)$ and $g_R(\mu')$ depends
only on $\mu$ and
$\mu'$ and not on $\Lambda$. Thus, being dimensionless, the 
$\beta$-function cannot depend on $\mu$ alone and is thus only a
function of $g_R$. This property is general for any renormalizable
theory: in the space of coupling constants, the $\beta$-function
is always a {\em local} function. Third, the $\beta$-function is
the function to be expanded in perturbation theory because it is
given by a true series in $g_R$ and not in $g_R \log\Lambda/\mu$.
This is clear for our example, Eq.~(\ref{betaordre1}), where there
is no logarithm, and can be proven formally by the following
argument. If we use Eq.~(\ref{fonctionfgr}) and Eq.~(\ref{beta}),
we find that:
\begin{equation}
\beta(g_R) =-{\frac{\partial f}{\partial y}(g_R,y)_\big|}_{y=1}.
\label{betaf}
\end{equation}
If $f$ is a double series in $g$ and in log($\mu/\mu'$),
\begin{equation}
f\big(g_R,\frac{\mu}{\mu'}\big)=\sum_{n,p}\alpha_{n,p} g_R^n 
\log^p\frac{\mu}{\mu'},
\label{doubleserie}
\end{equation}
it is clear from Eq.~(\ref{betaf}) that only terms with $p=1$
contribute to $\beta(g)$, with the logarithm replaced by $-1$.
Thus we immediately deduce from this argument and from
Eq.~(\ref{serief}) that:
\begin{equation}
\beta(g_R)=-\alpha g_R^2 -\beta g_R^3 -\gamma g_R^4 +O(g_R^5).
\label{beta_ordre_g4}
\end{equation}
It is easy to check that the first two coefficients, $-\alpha$ and
$-\beta$, are universal in the sense that for two different
theories, parametrized by $(g_R,\mu)$ and $(g_R',\mu)$, the two
$\beta$-functions have the same first two coefficients in their
expansions.

This method of computing the $\beta$-function also lets us by-pass
the strange way to calculate it that we have used in
Eqs.~(\ref{betag0}) and (\ref{betaordre1}) where we have first
expressed $g_R$ in terms of $g_0$ to calculate $\beta(g_R)$ as a
function of $g_0$ and then, by inversion of the series,
re-obtained a function of $g_R$. These two steps are {\em a
priori} dangerous because they both involve large logarithms.
Actually, they always cancel each other. This can be seen directly
for the example of Eq.~(\ref{relationordre3grg0}) and the 
reason for this cancellation comes from Eqs.~(\ref{betaf}) and
(\ref{doubleserie}), which shows that no inversion of series is
needed to calculate $\beta(g_R)$. There is no miracle here,
because only the behavior at
$y=\mu/\mu'=1$, which of course does not involve $\Lambda$,
matters.

Finally, we mention that the integration of the $\beta$-function at
$O(g_R^3)$ --- analogous to a two-loop result in QFT --- leads to
an implicit equation for $g_R'$ that generalizes
Eq.~(\ref{grgrprimeintegre}):
\begin{equation}
\frac{1}{g_R'}- \frac{1}{g_R} +\frac{\beta}{\alpha}
\log\big(\frac{g_R}{g_R'} \frac{\alpha +\beta g_R'}{\alpha +\beta
g_R} \big)=
\alpha\log\frac{\mu'}{\mu}.
\label{two_loop_int}
\end{equation}
There is no simple solution of this transcendental equation. It is
however possible to obtain an iterative solution that is valid if
the $O(g_R^3)$ term is small compared with the
$O(g_R^2)$ one, that is, if $g_R \beta/\alpha\ll 1$. It is
obtained by replacing $g_R'$ in the third term of
Eq.~(\ref{two_loop_int}) by its expression obtained to order
$g_R^2$, Eq.~(\ref{grgrprimeintegre}):
\begin{equation}
g_R'= \frac{g_R}{1-\alpha g_R \log\frac{\mu}{\mu'}
+\frac{\beta}{\alpha} g_R\log\big(1-\alpha g_R
\log\frac{\mu}{\mu'} \big)}.
\label{this}
\end{equation}
It is easy to check that Eq.~(\ref{this}) resums exactly all the
leading and sub-leading logarithms of the perturbation expansion
Eq.~(\ref{Ftot_ren}). Note that contrarily to the one-loop
result, Eq.~(\ref{grgrprimeintegre}), which resums only the
leading logarithms, the exact expression
in Eq.~(\ref{two_loop_int}) contributes also to the sub-sub-leading
logarithms as well as the sub-sub-sub-leading ones and so on and
so forth.

\section{Summary}

(1) The long way of renormalization starts with a theory
depending on only one parameter $g_0$, which is the small
parameter in which perturbation series are expanded. In particle
physics, this parameter is in general a coupling constant like an
electric charge involved in a Hamiltonian (more precisely the fine
structure constant for electrodynamics). This parameter is also
the first order contribution of a physical quantity $F$. In
particle/statistical physics,
$F$ is a Green/correlation function. The first order of
perturbation theory neglects fluctuations --- quantum or
statistical --- and thus corresponds to the classical/mean field
approximation. The parameter $g_0$ also is to this order a
measurable quantity because it is given by a Green function. Thus,
it is natural to interpret it as the unique and physical coupling
constant of the problem. If, as we suppose in the following, $g_0$
is dimensionless, so is $F$. Moreover, if $x$ is dimensional ---
it represents momenta in QFT --- it is natural that $F$ does not 
depend on it as is found in the classical theory, that is, at
first order of the perturbation expansion.

(2) If $F$ does depend on $x$, as we suppose it does at second
order of perturbation theory, it must depend on another
dimensional parameter, $\Lambda$, through the ratio of $x$ and 
$\Lambda$. If we have not included this parameter from the
beginning in the model, the $x$-dependent terms are either
vanishing, which is what happens at first order, or infinite as
they are at second and higher orders. This is the very origin of
divergences (from the technical point of view).

(3) These divergences require that we regularize $F$. This
requirement, in turn, requires the introduction of the scale
$\Lambda$ that was missing. In the context of field theory, the
divergences occur in Feynman diagrams for high momenta, that is,
at short distances. The cut-off
$\Lambda$ suppresses the fluctuations at short distances compared
with $\Lambda^{-1}$. In statistical physics, this scale, although
 introduced for formal reasons, has a natural
interpretation because the theories are always 
effective theories built at a given microscopic scale. It
corresponds in general to the range of interaction of the
constituents of the model, for example, a lattice spacing for
spins, the average intermolecular distance for fluids. In
particle physics, things are less simple. At least
psychologically. It was indeed natural in the early days of
quantum electrodynamics to think that this theory was fundamental,
that is, not derived from a more fundamental theory. More
precisely, it was believed that QED had to be mathematically
internally consistent, even if in the real world new physics had to
occur at higher energies. Thus, the regulator scale was introduced
only as a trick to perform intermediate calculations. The limit
$\Lambda\to\infty$ was supposed to be the right way to eliminate
this unwanted scale, which anyway seemed to have no interpretation.
We shall see in the following that the community now interprets 
the renormalization process differently.

(4) Once the theory is regularized, $F$ can be a non-trivial
function of $x$. The price is that different values
of $x$ now correspond to different values of the coupling constant
(defined as the values of $F$ for these $x$). Actually, it does
no longer make sense to speak of a coupling constant in itself.
The only meaningful concept is the pair $(\mu,g_R(\mu))$
of coupling constants at a given scale. The relevant question
now is, ``What are the physical reasons in particle/statistical
physics that make the coupling constants depend on the scale while
they are constants in the classical/mean field approximation?'' As 
mentioned, for particle physics, the answer is the existence of
new quantum fluctuations corresponding to the possibility of
creating (and annihilating) particles at energies higher than $m
c^2$. What was scale independent in the classical theory becomes
scale dependent in the quantum theory because, as the available
energy increases, more and more particles can be created. The
pairs of (virtual) particles surrounding an electron are polarized
by its presence and thus screen its charge. As a consequence, the
charge of an electron depends on the distance (or equivalently the
energy) at which it is probed, at least for distances smaller than
the Compton wavelength. 

Note that the energy scale $m c^2$ should not be confused with
the cut-off scale
$\Lambda$. $m c^2$ is the energy scale above which quantum
fluctuations start to play a significant role while
$\Lambda$ is the scale where they are cut-off. Thus, although the
Compton wave length is a short distance scale for the classical
theory, it is a long distance scale for QFT, the short one being
$\Lambda^{-1}$. There are thus three domains of length scales in
QFT: above the Compton wave length where the theory behaves
classically (up to small quantum corrections coming from high
energy virtual processes), between the Compton wave length and
the cut-off scale $\Lambda^{-1}$ where the relativistic and
quantum fluctuations play a great role, and below $\Lambda^{-1}$
where a new, more fundamental theory has to be
invoked.\cite{lepage89} In statistical physics, the analogue of
the Compton wave length is the correlation length which is a
measure of the distance at which two microscopic constituents of
the system are able to
influence each other through thermal fluctuations.\cite{foot11}
 For the Ising model for instance, the
correlation length away from the critical point 
 is the order of the lattice spacing and the
corrections to the mean-field approximation due to fluctuations
are small. Unlike particle physics where the masses and therefore
the Compton wavelengths
are fixed, the correlation lengths in statistical mechanics can
be tuned by varying the temperature. Near the critical temperature
where the phase transition
takes place, the correlation length becomes extremely large 
and fluctuations on all length scales between the microscopic
scale of order
$\Lambda^{-1}$, a lattice spacing,
and the correlation length add up to modify the mean-field
behavior (see Refs.~\onlinecite{raposo91}, \onlinecite{stanley99} 
and also Ref.~\onlinecite{tobochnik01} for a 
bibliography on this subject). We see here a key to the
relevance of renormalization: two very different scales must exist
between which a non-trivial dynamics (quantum or statistical in
our examples) can develop. This situation is {\em a priori} rather
unnatural as can be seen for phase transitions,
where a fine tuning of temperature must be implemented to obtain
correlation lengths much larger than the microscopic scale. Most
of the time, physical systems have an intrinsic scale (of time,
energy, length, etc) and all the other relevant scales of the
problem are of the same order. All phenomena occurring at very
different scales are thus almost completely suppressed. The existence
of a unique relevant scale is one of the reasons why renormalization is not
necessary in most physical theories. In QFT it is mandatory
because the masses of the known particles are much smaller
than a hypothetical cut-off scale $\Lambda$, still to be
discovered, where new physics should take place. This is a rather
unnatural situation, because, contrary to phase transitions, there
is no analogue of a temperature that could be fine-tuned to
create a large splitting of energy, that is, mass, scales. The
question of naturalness of the models we have at present in
particle physics is still largely open, although there has been
much effort in this direction using supersymmetry.

(5) The classical theory is valid down to the Compton/correlation
length, but cannot be continued naively beyond this scale;
otherwise, when mixed with the quantum formalism, it produces
divergences. Actually, it is known in QFT that the fields should
be considered as distributions and not as ordinary functions.
The need for considering distributions comes from the non-trivial structure of the
theory at very short length scale where fluctuations are very
important. At short distances, functions
are not sufficient to describe the field state, which is not smooth
but rough, and distributions are necessary. Renormalizing the
theory consists actually in building, order by order, the correct
``distributional continuation'' of the classical theory. The
fluctuations are then correctly taken into account and depend on
the scale at which the theory is probed: this non-trivial scale
dependence can only be taken into account theoretically through
the dependence of the (analogue of the) function
$F$ with $x$ and thus of the coupling with the scale
$\mu$.

(6) If the theory is perturbatively renormalizable, the pairs
$(\mu,g(\mu))$ form an equivalence class of parametrizations of
the theory. The change of parametrization from $(\mu,g(\mu))$ to
$(\mu',g(\mu'))$, called a renormalization group transformation,
is then performed by a law which is self-similar, that is, such
that it can be iterated several times while being
form-invariant.\cite{kovalev99,shirkov01} This law is obtained by
the integration of 
\begin{equation}
\beta(g_R)=\mu\frac{\partial
g_R}{\partial\mu}\bigg|_{g_0,\Lambda}.
\label{this1}
\end{equation}
This function has a true perturbation expansion in terms of $g_R$
unlike the perturbative relation between
$g_R(\mu)$ and $g_R(\mu')$ which involves logarithms of $\mu/\mu'$
that can be large. The integration of Eq.~(\ref{this1}) 
partially resums the perturbation series and is thus
semi-non-perturbative even if
$\beta(g_R)$ has been calculated perturbatively. The self-similar
nature of the group law is encoded in the fact that $\beta(g_R)$
is independent of
$\Lambda$.\cite{lebellac91}

In particle physics, the $\beta$-function gives the evolution of the strength of the interaction as the energy
at which it is probed varies and the integration of the $\beta$-function resums partially the
perturbation expansion. First, as the
energy increases, the coupling constant can decrease and eventually
vanish. This is what happens when $\alpha>0$ in
Eqs.~(\ref{betaordre1}) and ({\ref{grgrprimeintegre}). In this
case, the particles almost cease to interact at very high energies
or equivalently when they are very close to each other. The theory
is then said to be asymptotically free in the ultraviolet
domain.\cite{collins84,lebellac91} Reciprocally, at low energies
the coupling increases and perturbation theory can no longer be
trusted. A possible scenario is that bound states are created at a
sufficiently low energy scale so that the perturbation approach
has to be reconsidered in this domain to take into account these
new elementary excitations. Non-abelian gauge theories are the
only known theories in four space-time dimensions that are
ultraviolet free, and it is widely believed that quantum
chromodynamics --- which is such a theory --- explains quark
confinement. The other important behavior of the scale
dependence of the coupling constant is obtained
for $\alpha<0$ in which case it increases at high energies. This 
corresponds for instance to quantum electrodynamics. For this kind of theory,
the dramatic increase of the coupling at high energies is supposed
to be a signal that the theory ceases to be valid beyond a certain
energy range and that new physics, governed by an asymptotically
free theory (like the standard model of electro-weak
interactions) has to take place at short distances. 

(7) Renormalizability, or its non-perturbative equivalent,
self-similarity, ensures that although the theory is initially
formulated at the scale $\Lambda$, this scale together with $g_0$
can be entirely eliminated for another scale better adapted to the
physics we study. If the theory was solved exactly, it would make
no difference which parametrization we used.
However, in perturbation theory, this renormalization lets us
avoid calculating small numbers as differences of very large
ones. It would indeed be very unpleasant, and
actually meaningless, to calculate energies of order 100\,GeV, for
instance --- the scale
$\mu$ of our analysis --- in terms of energies of order of the 
Planck scale $\simeq 10^{19}$\,GeV, the analogue of the scale
$\Lambda$. In a renormalizable theory, the possibility to
perturbatively eliminate the large scale has a very deep
meaning: it is the signature that the physics is short
distance insensitive or equivalently that there is a 
decoupling of the physics at different scales. The only memory of
the short distance scale lies in the initial conditions of the
renormalization group flow, not in the flow itself: the
$\beta$-function does not depend on $\Lambda$.We again
emphasize that, usually, the decoupling of the physics at very
different scales is trivially related to the existence of a
typical scale such that the influence of all phenomena occurring at
different scales is almost completely suppressed. Here, the
decoupling is much more subtle because there is no typical length
in the whole domain of length scales that are very small compared
with the Compton wave length and very large compared with 
$\Lambda^{-1}$. Because interactions among particles correspond to
non-linearities in the theories, we could naively believe that all
scales interact with each others --- which is true --- so that
calculating, for instance, the low energy behavior of the theory
would require the detailed calculation of all interactions
occurring at higher energies. Needless to say that in a field
theory, involving infinitely many degrees of freedom --- the value
of the field at each point --- such a calculation would be 
hopeless, apart from exactly solvable models. Fortunately, such a calculation
is not necessary for physical quantities that can be calculated
from renormalizable couplings only. Starting at very high
energies, typically
$\Lambda$, where all coupling constants are naturally of order 1,
the renormalization group flow drives almost all of them to zero,
leaving only, at low energies, the renormalizable couplings. This
is the interpretation of non-renormalizable couplings. They are
not terrible monsters that should be forgotten as was believed in
the early days of QFT. They are simply couplings that the RG flow
eliminates at low energies. If we are lucky, the renormalizable
couplings become rather small after their RG evolution between
$\Lambda$ and the scale $\mu$ at which we work, and perturbation
theory is valid at this scale.

We see here the phenomenon of universality: among
the infinitely many coupling constants that are {\it a priori}
necessary to encode the dynamics of the infinitely many degrees of
freedom of the theory, only a few ones are finally
relevant.\cite{bagnuls01} All the others are washed out at large
distances. This is the reason why, perturbatively, it is not
possible to keep these couplings finite at large distance, and
it is necessary to set them to zero.\cite{foot12} The simplest
non-trivial example of universality is given by the law of large
numbers (the central limit theorem) which is crucial
in statistical mechanics.\cite{raposo91} In systems where it can
be applied, all the details of the underlying probability
distribution of the constituents of the system are irrelevant for
the cooperative phenomena which are governed by a gaussian
probability distribution.\cite{jona-lasinio00} This drastic
reduction of complexity is precisely what is necessary for physics
because it lets us build effective theories in which only a
few couplings are kept.\cite{lepage89} Renormalizability in
statistical field theory is one of the non-trivial generalizations
of the central limit theorem.

(8) The cut-off $\Lambda$, first introduced as a mathematical
trick to regularize integrals, has actually a deep physical
meaning: it is the scale beyond which new physics occur and below
which the model we study is a good effective description of
the physics. In general, it involves only the renormalizable
couplings and thus cannot pretend to be an exact description of
the physics at all scales. However, if $\Lambda$ is very large
compared with the energy scale in which we are interested, all
non-renormalizable couplings are highly suppressed and the
effective model, retaining only renormalizable couplings, is valid
and accurate (the Wilson RG formalism is well suited to this
study, see Refs.~\onlinecite{bagnuls01} and
\onlinecite{tetradis94}). In some models --- the asymptotically
free ones --- it is possible to formally take the limit $\Lambda\to\infty$
both perturbatively and non-perturbatively, and there is therefore
no reason to invoke a more fundamental theory taking over at a
finite (but large) $\Lambda$. Let
us emphasize here several interesting points.

\begin{enumerate}[(i)]

\item For a theory corresponding to the pair
$(\mu,g_R(\mu))$, the limit $\Lambda\to\infty$ must be taken
within the equivalence class of parametrizations to which 
$(\mu,g_R(\mu))$
belongs.\cite{nonperb} A
divergent non-regularized perturbation expansion consists in
taking $\Lambda=\infty$ while keeping $g_0$ finite. From this
viewpoint, the origin of the divergences is that the pair
$(\Lambda=\infty,g_0)$ does not belong to any equivalence class of
a sensible theory. Perturbative renormalization consists in
computing $g_0$ as a formal powers series in $g_R$ (at finite
$\Lambda$), so that $(\Lambda,g_0)$ corresponds to a mathematically
consistent theory; we then take the limit $\Lambda\to\infty$.

\item Because of universality, it is physically impossible to know from low
energy data if $\Lambda$ is very large or truly infinite.

\item Although mathematically consistent, it seems unnatural to
reverse the RG process while keeping only the renormalizable
couplings and thus to imagine that even at asymptotically high
energies, Nature has used only the couplings that we are able to
detect at low energies. It seems more natural that a fundamental
theory does not suffer from renormalization problems. String
theory is a possible candidate.\cite{schmidhuber97}

\end{enumerate}

To conclude, we see that although the renormalization procedure has
not evolved much these last thirty years, our interpretation of 
renormalization has drastically changed\cite{lepage89}: the
renormalized theory was assumed to be fundamental, while it is now
believed to be only an effective one; $\Lambda$ was interpreted as
an artificial parameter that was only useful in intermediate
calculations, while we now believe that it corresponds to a
fundamental scale where new physics occurs; non-renormalizable
couplings were thought to be forbidden, while they are now
interpreted as the remnants of interaction terms in a more
fundamental theory. Renormalization group is now seen as
an efficient tool to build effective low energy theories when
large fluctuations occur between two very different scales that
change qualitatively and quantitatively the physics.

\begin{acknowledgments}
The author thanks S.\ Dusuel and B.\ Dou\c{c}ot for many interesting
discussions and remarks. He also thanks J.\ Bartlett for his help
in the choice of title of this article, and L.\ Canet, A.\
Elkharrat, E.\ Huguet, A.\ Laverne, P.\ Lecheminant, R.\ Mosseri,
D.\ Mouhanna, N.\ Wschebor and M.\ Tissier for a careful reading
of the manuscript and many suggestions. He also thanks all the
students of the Ninth Vietnam School of Physics (Hu\'e, January
2003) for having worked hard on this article and for their
suggestions. Laboratoire de Physique Th\'eorique et Hautes Energies
is a Laboratoire associ\'e au CNRS: UMR 7589.
\end{acknowledgments}

\appendix
\section{Toy models for renormalizable and non-renormalizable
perturbation expansions} We give an example of a non-renormalizable
theory and of a theory which needs two couplings to be
renormalized. 
Let us suppose that 
\begin{equation}
F_{1,\Lambda}(x)= \alpha \!\int_1^\Lambda dt \frac{t}{t+x}
\label{exemple2}
\end{equation}
which, unlike the example of Eq.~(\ref{exempleregularise}), is
linearly divergent. To renormalize this function, we have to impose
a prescription at one point, and we choose:
\begin{equation}
F_{\Lambda}(0)= g_R.
\label{pres_exemple2}
\end{equation}
Note that it was not possible in the example 
of Eq.~(\ref{exempleregularise}) to take $\mu=0$, because this
choice would have lead to a divergence of the integral at the lower
bound. In Eq.~(\ref{exemple2}) taking $\mu=0$ is possible
because the lower bound of the integral
is non-vanishing and actually plays somewhat the role of a
non-vanishing $\mu$. We have,
\begin{equation}
\delta_2 g= -\alpha g_R^2 \int_1^\Lambda dt,
\end{equation}
so that:
\begin{equation}
F_{\Lambda}(x)=g_R - \alpha g_R^2\, x\! \int_1^\Lambda
\frac{dt}{t+x},
\end{equation}
which is still (logarithmically) divergent for all $x\ne 0$. The
difference between the two examples given in Eqs.(\ref{exempleregularise}) and (\ref{exemple2}) is that in the last one, once
the linear divergence has been subtracted, the logarithmic
sub-divergence remains. Subtracting it would require us to impose a
second prescription that would define a new coupling constant. In
the absence of this second coupling constant, the logarithmic
divergence cannot be subtracted and the model is
non-renormalizable.

Let us examine how a second coupling constant could solve the
problem. Generically, this second coupling, which we call
$\lambda_0$, already contributes at first order. We take as an
example:
\begin{equation}
F_{\Lambda}(x)=g_0 +\lambda_0 x + \alpha g_0^2 \int_1^\Lambda dt 
\frac{t}{t+x}+O(g_0^3).
\label{exemple3}
\end{equation}
Let us take as renormalization prescriptions,
\begin{equation}
\frac{\partial F_{\Lambda}}{\partial x} (x=0)=\lambda_R,
\end{equation}
in addition to Eq.~(\ref{pres_exemple2}). We obtain at first order
that $g_0=g_R +O(g_R^2)$ and $\lambda_0=\lambda_R +O(g_R^2)$ and
at second order:
\begin{eqnarray}
\delta_2 g&=& -\alpha g_R^2
\int_1^\Lambda dt\\
\delta_2 \lambda&=& \alpha g_R^2 \int_1^\Lambda
\frac{dt}{t}.
\end{eqnarray}
If we substitute these expressions in Eq.~(\ref{exemple3}), we
find:
\begin{equation}
F_{\Lambda}(x)=g_R +\lambda_R x + \alpha g_R^2 x^2\int_1^\Lambda
\frac{dt}{t(t+x)}+O(g_R^3).
\label{a8}
\end{equation} Obviously, this expression converges when
$\Lambda\to \infty$. The two renormalization prescriptions let us
subtract the linear divergence as well as the logarithmic
sub-divergence. We emphasize that in the previous example we 
only eliminated the second divergence at order $g_0^2$. At higher
orders, there are two ways a theory can behave, characterized by
two different renormalizability properties. The first one is that
all divergences can be removed to all orders by renormalizing
only the two couplings $g_0$ and $\lambda_0$. A variant of this
possibility is that a third coupling --- or a finite number of new
couplings --- turns out to be necessary and sufficient to remove
the divergences. In this case, the model is renormalizable at the price of introducing all
the necessary couplings. The
second possibility is that the new interaction term, which has
induced the existence of the $\lambda_0$ term in $F$, generates 
new divergences at
higher orders. In this case, new interaction terms (and coupling
constants) are required to remove the new divergences. These
new terms can themselves generate new
divergences at even higher orders, which require new couplings
to be removed and so on and so forth. In this case, infinitely
many interaction terms are necessary to remove the divergences
and the model is perturbatively non-renormalizable.

\section{DERIVATION OF EQ.(22)}
Let us show that it is always possible to make the choice used in
Eq.~(\ref{condition1}). Due to condition
(\ref{conditionpremierordre}), we have generally:
\begin{equation}
F_{1,\Lambda}^{s}(x)-F_{1,\Lambda}^{s}(\mu)\ \to\ g'_1(x,\mu),\
\Lambda\to\infty
\label{b1}
\end{equation}
where the limit $g'_1$ is a well defined function satisfying
$g'_1(x,\mu)=-g'_1(\mu,x)$. If we first evaluate 
$F_{1,\Lambda}^{s}(x)$ in Eq.~(\ref{b1}) at
$x=1$ for instance and at $\mu=1$ and subtract them, we
obtain that $g_1'$ has the form:
\begin{equation}
g'_1(x,\mu)=g_1(x)-g_1(\mu),
\end{equation}
namely, a combination of the same function of $x$ and $\mu$. Then,
by redefining $F_{1,\Lambda}^{s}$: $F_{1,\Lambda}^{s}\to
F_{1,\Lambda}^{s}-g_1$ we satisfy Eq.~(\ref{condition1}). Note
that the previous choice of singular part is not necessary and is
only convenient. 

\section{Logarithmic divergences in renormalizable theories with dimensionless couplings}
We prove  for renormalizable theories with dimensionless
couplings that $F_{1,\Lambda}^s(x)$ must be a logarithm. If we use
Eq.~(\ref{solutionf1}), dimensional analysis, and the freedom to
choose the regular part of 
$F_{1,\Lambda}$, we have,
\begin{equation}
F_{1,\Lambda}^s(x)=f\big(\frac{x}{\Lambda}\big)=f_1(\Lambda) +
r(x).
\label{eq1}
\end{equation}
Note that in full generality, the regular part we add to
$f_1(\Lambda)$ could depend on $\Lambda$: $r_\Lambda(x)$. However,
because it is regular, we can choose to add only the
$\Lambda$-independent function corresponding to the
$\Lambda\to\infty$ limit of $r_\Lambda$: $r(x)=r_\infty(x)$. If
we differentiate Eq.(\ref{eq1}) with respect to $x$ and then take
$x=1$ and
$\Lambda=1/y$, we obtain,
\begin{equation}
f'(y)=\frac{r'(1)}{y},
\end{equation}
and thus
\begin{equation}
f(y)= -\alpha \log y,
\label{c3}
\end{equation}
where the minus sign has been written for convenience. From
(\ref{eq1}) and (\ref{c3}) we conclude that
$f(x)=r(x)=-f_1(x)$ and that
\begin{equation}
F_{1,\Lambda}^s(x)=f\big(\frac{x}{\Lambda}\big)=
f(x)-f(\Lambda)=\alpha
\log\frac{\Lambda}{x}.
\label{eq2}
\end{equation}

\section{Renormalization group improved expansion}
We show how to derive Eq.~(\ref{equation_G}).
Consider the definition of $F^{\rm imp}$:
\begin{equation}
F^{\rm imp}(x,g_0,\Lambda)=g_0 + \alpha g_0^2
\log\big(\frac{\Lambda}{x}\big) + g_0^3
G\big(\frac{\Lambda}{x}\big)+ O(g_0^4)
\label{fimp_appendice}
\end{equation}
We can calculate $g_R$ and $g_R'$ from their definitions (where
$F^{\rm imp}$ is used instead of $F$) and from
Eq.~(\ref{fimp_appendice}):
\begin{eqnarray}
g_R&=&g_0 +\alpha g_0^2 \log\big(\frac{\Lambda}{\mu}\big) +
g_0^3 G\big(\frac{\Lambda}{\mu}\big)+ O(g_0^4)\\
g_R'&=&g_0 +\alpha g_0^2 \log\big(\frac{\Lambda}{\mu'}\big) +
g_0^3 G\big(\frac{\Lambda}{\mu'}\big) + O(g_0^4).
\label{grpordre3}
\end{eqnarray}
If we invert the series $g_R=g_R(g_0)$ of Eq.~(\ref{grpordre3}), we
obtain:
\begin{equation}
g_0=g_R-\alpha g_R^2\log\big(\frac{\Lambda}{\mu}\big)+ 2
\alpha^2 g_R^3 \log^2\big(\frac{\Lambda}{\mu}\big)
- g_R^3 G\big(\frac{\Lambda}{\mu}\big)
+O(g_R^4).
\end{equation}
Thus, substituting this expression for $g_0$ in $g_R'=g_R'(g_0)$,
Eq.~(\ref{grpordre3}), we obtain:
\be
\begin{array}{ll}
g_R'=g_R  +&\displaystyle{\hspace{-0.1cm}\alpha g_R^2 \log\left(\frac{\mu}{\mu'}\right) + 
g_R^3\left( 2\alpha^2\left( \log^2\left(\frac{\Lambda}{\mu}\right)-\right.\right.} \\
&\hspace{-0.3cm}\displaystyle{-\left.\left.\log\left(\frac{\Lambda}{\mu} \right)\log\left(\frac{\Lambda}{\mu'}\right)\right) + G\left(\frac{\Lambda}{\mu'}\right)- 
G\left(\frac{\Lambda}{\mu}\right) \right)}.
\end{array}
\ee
The group law is obeyed at this order if the relation between
$g_R'$ and $g_R$ is of the same form as the one between
$g_R$ and $g_0$, Eq.~(\ref{grpordre3}). This condition requires:
\begin{equation}
g_R'=g_R +\alpha g_R^2 \log\big(\frac{\mu}{\mu'}\big) + g_R^3
G\big(\frac{\mu}{\mu'}\big)+ O(g_R^4),
\end{equation}
and thus:
\begin{equation}
2 \alpha^2\log\frac{\Lambda}{\mu} \log\frac{\mu'}{\mu} +
G\big(\frac{\Lambda}{\mu'}\big)-
G\big(\frac{\Lambda}{\mu}\big)= G\big(\frac{\mu}{\mu'}\big).
\label{equaG}
\end{equation}
By differentiating this relation with respect to $\Lambda$ and by taking $\Lambda=\mu$, we find, setting $x=\mu/\mu'$:
\begin{equation}
G'(x)=2 \alpha^2 \frac{\log x}{x} + \frac{G'(1)}{x}.
\end{equation}
If we take into account that $G(1)=0$, we find
by integration:
\begin{equation}
G(x)=\alpha^2 \log^2x +\beta \log x,
\end{equation}
where $\beta$ is arbitrary.

\section{The renormalization group applied to a differential equation}
\label{appendix_E}
We show how the renormalization program can be
implemented for the example of the differential equation
Eq.~(\ref{trivial_equa_diff}) whose exact solution is
\begin{equation}
y(t)=f(r_0,t-t_0)=r_0 e^{\epsilon(t-t_0)}.
\end{equation}

\begin{figure}[htbp] 
\begin{center}
\includegraphics[width=.9\linewidth,origin=tl]{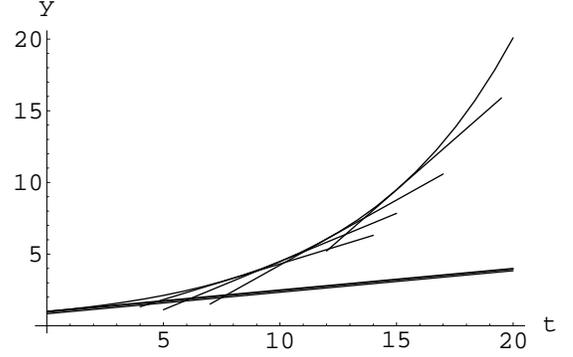}\hfill%
\end{center}
\caption{The curve $y(t)$ as a function of $t$. The (thick) lower line is the approximation of order $\epsilon$, see Eq.(\ref{approx_ordre_eps}). 
The other lines represent the (field of) tangent vectors to the curve  --- the envelope ---  given by the $\beta$-function, Eq.(\ref{valeurbeta}).}
\label{enveloppe}
\end{figure}

In perturbation theory, we find:
\begin{equation}
y(t)= r_0 (1+\epsilon(t-t_0) +\frac{\epsilon^2}{2}(t-t_0)^2
+\dots).
\label{approx_ordre_eps}
\end{equation}
At order $\epsilon^0$, $y(t)$ is constant and finite, whereas, at
any higher order in $\epsilon$, a divergence occurs for
$t_0\to-\infty$. This divergence arises of course from the fact
that the expansion turns out to be in powers of 
$\epsilon(t-t_0)$ and not in powers of $\epsilon$ alone (the
secular problem). Thus, as shown in Fig.\ref{enveloppe}, the
approximation of order
$O(\epsilon)$ becomes worse and worse as
$t$ increases. A renormalization prescription consists here in
imposing that for a finite $\tau$:
\begin{equation}
y(\tau)=r_\tau.
\label{tauprescription}
\end{equation}
If we perform the calculation to order $\epsilon$, we find to
first order:
\begin{equation}
r_\tau= r_0\big(1+\epsilon(\tau-t_0)\big) +O(\epsilon^2),
\label{rtau}
\end{equation}
and thus, as expected:
\begin{equation}
y(t)=r_\tau\big(1+\epsilon(t-\tau)\big) +O(\epsilon^2).
\label{yrtau}
\end{equation}
The theory is perturbatively renormalizable at this order because by imposing a
single renormalization prescription, it is possible to completely
eliminate
$t_0$ and $r_0$.
Let us define the $\beta$-function for $r_\tau$ by:
\begin{equation}
\beta(r_\tau) = \frac{\partial
r_\tau}{\partial\tau}\bigg|_{r_0,t_0}=\frac{\partial f}{\partial
\zeta}(r_\tau,\zeta)\bigg|_{\zeta=0}.
\label{betar}
\end{equation}
We find:
\begin{equation}
\beta(r_\tau) = \epsilon r_\tau +O(\epsilon^2).
\label{valeurbeta}
\end{equation}

It is very instructive to perform this calculation at higher
orders because we then find that the $O(\epsilon)$ result of
Eq.~(\ref{valeurbeta}) is exact (this result is trivially shown
using the second equality of Eq.~(\ref{betar})). Thus, there is no
$O(\epsilon^2)$ corrections to $\beta(r_\tau)$. This result means that, in this example, there is no subleading terms
such as
$\epsilon^n (t-t_0)^p$ with $p<n$ in the perturbation expansion.

Clearly, the $\beta$-function gives the tangent to the curve
$y(t)$. Equation~(\ref{valeurbeta}) shows that contrarily to
$y(t)$, the
$\beta$-function has a true $\epsilon$-expansion (involving only
one term
in this example). 
This result is reminiscent of what we have already observed
in our general discussion, see Eqs.~(\ref{serief}) and
(\ref{beta_ordre_g4}). Of course, this example is too simple
because using the $\beta$-function leads to the same 
differential equation for $r_\tau$ as the one for $y(t)$ that we
started with, Eq.~(\ref{trivial_equa_diff}): the RG does not help
us to solve ordinary differential equations. 
However, although mathematically trivial, our analysis
shows that perturbation theory should not be used for large
$t-t_0$, but that it is perfectly valid for infinitesimal time
steps, see Fig.~\ref{enveloppe}. It also shows that
the higher order terms of the perturbation expansion are
completely analogous to the series of the leading logarithms we
have previously encountered: they are entirely determined by the
$O(\epsilon)$ term together with self-similarity (encoded in the
$\beta$-function). Note finally that for partial differential
equations (PDE) that describe the dynamics of infinitely many
degrees of freedom (as in field theory), the RG techniques do not
let us reconstruct the PDE from the first orders of perturbation
theory. The $\beta$-functions lead to ordinary
differential equations, the integration of which let us improve the
perturbation computation of several quantities thanks to a partial
resummation of the perturbation
expansion.\cite{shirkov01,kovalev99}


\begin{thebibliography}{10}
\bibitem{foot1}Some phase transitions are triggered by quantum
fluctuations. This subtlety plays no role in what follows.
\bibitem{foot2}The short distance physics in statistical systems
is given by Hamiltonians describing, for instance, interactions
among magnetic ions or molecules of a fluid.
\bibitem{foot3} Actually, the analogue of the function in
Eq.~(\ref{developpement}) would be a correlation function of four
density or spin fields taken at four different points. These 
functions are not easily measurable and thus $g_0$ does not have in
general an intuitive meaning in this case. Because this subtlety 
plays no role in our discussion, we ignore this
difficulty in the following.
\bibitem{foot4}Actually for QED it is a
four-dimensional integral over four-momenta and the integrand is a
product of propagators.
\bibitem{foot5}It is nontrivial to prove in full generality that
the results obtained after renormalization are independent of the
regularization scheme. However, it is easy to grasp the idea behind
it. Because renormalization consists in eliminating parameters like
$g_0$ and in replacing them by measurable couplings like $g_R$, the
renormalized quantities like $F(x)$ are finally expressed only in
terms of physical quantities that are independent of the
regularization scheme.\cite{gosdzinsky91}
\bibitem{foot6}Note that the (renormalized) series in
$g_R$ can themselves be non-convergent. Most of the time
they are at best asymptotic. In some cases they can be resummed
using Borel transform and Pad\'e approximants.
\bibitem{foot7}In QFT, it is in general also necessary to change
the normalization of the analogue of the function $F$ --- the Green
functions --- by a factor that diverges in the limit
$\Lambda\to\infty$. This procedure is known as field
renormalization.
\bibitem{foot8}Let us emphasize that there is a subtlety if
dimensional regularization is used. Actually, this
regularization also introduces a dimensional parameter $\lambda$,
which is not directly a regulator as is the cut-off $\Lambda$ in
the integral of Eq.~(\ref{exempleregularise}). The analogue of
$\Lambda$ in this regularization is given by
$\Lambda =\lambda\exp(1/\epsilon)$, where $\epsilon=4-D$ and $D$
is the spatial dimension. It often is convenient to take
$\lambda\sim\mu$. We mention that dimensional regularization
kills all non-logarithmic divergences.\cite{mitra98}
\bibitem{foot9}The elements of the group are the functions:
$g_t=f(\cdot\,,t)$ for $t\in \nbR$. They transform an initial
condition $r_0$ into the solution at a later time interval $t$
of the differential equation we consider (see Eq.~(\ref{def_fonction_f})
in our example): $g_t(r_0)=f(r_0,t)$. 
The composition law is thus $g_{t'}.g_{t}=f\left(f(\cdot\,,t),t'\right)$.
It obeys trivially the identity:  $g_{t'}.g_{t}=g_{t+t'}$ which is nothing
but Eq.(\ref{loi_de_groupe}).
The identity is $g_{t=0}$ and the inverse is $g_{-t}$. The law is associative
as it should for a group.
\bibitem{foot10}If we had not omitted in Eq. (\ref{eqfoot10})
 the finite parts, we would have found 
$F(x=\Lambda)=g_0+a g_0^2 +b g_0^3+\dots$ Thus
$g_0$ is in general not associated exactly with the scale
$\Lambda$, but with $\Lambda$ up to a factor of order unity.
\bibitem{foot11}It is quite similar to the
Compton wavelength of the pion which is the typical range of the
nuclear force between hadrons like protons and nucleons.
\bibitem{foot12}More precisely, working with non-renormalizable
couplings would require
us to fine-tune infinitely many of them at scale
$\Lambda^{-1}$ to unnatural values. Most of the time, such a
finely-tuned model is no longer predictive.
\bibitem{nonperb}Non-perturbatively, the existence of the limit
$\Lambda\to\infty$ is more subtle than perturbatively. The
renormalization group flow must be controlled in this limit and
this is achieved if non-perturbatively
$g_0$ has a finite limit, that is, if there exists an ultraviolet
fixed point of the RG flow. The case $g_0\to 0$ when
$\Lambda\to\infty$, corresponds to asymptotically free theories,
that is, in four space-time dimensions, to non-abelian gauge
theories.
\end{thebibliography}
\end{document}